\documentclass[12pt,amsbsy]{article}
\newcommand{\beq}{\begin{equation}}
\newcommand{\eeq}{\end{equation}}
\newcommand{\beqn}{\begin{eqnarray}}
\newcommand{\eeqn}{\end{eqnarray}}

\newcommand{\bJ}{\mathbf{J}}

\newcommand{\be}{\mbox{${\beta}$}}

\newcommand{\de}{\mbox{${\delta}$}}

\begin{document}

\begin{titlepage}

\vspace{1cm}

\begin{center}
{\bf \large Spectrum of quantized black hole,

\vspace{2mm}

correspondence principle, and holographic bound}
\end{center}

\begin{center}
I.B. Khriplovich \footnote{khriplovich@inp.nsk.su}
\end{center}
\begin{center}
Budker Institute of Nuclear Physics\\
630090 Novosibirsk, Russia,\\
and Novosibirsk University
\end{center}

\bigskip

\begin{abstract}
An equidistant spectrum of the horizon area of a quantized black
hole does not follow from the correspondence principle or from
general statistical arguments. On the other hand, such a spectrum
obtained in loop quantum gravity (LQG) either does not comply with
the holographic bound, or demands a special choice of the
Barbero-Immirzi parameter for the horizon surface, distinct from
its value for other quantized surfaces. The problem of
distinguishability of edges in LQG is discussed, with the
following conclusion. Only under the assumption of partial
distinguishability of the edges, the microcanonical entropy of a
black hole can be made both proportional to the horizon area and
satisfying the holographic bound.
\end{abstract}

\vspace{8cm}

\end{titlepage}

{\bf 1.} The idea of quantizing the horizon area of black holes
was put forward many years ago by Bekenstein in the pioneering
article~\cite{bek}. It was based on the intriguing observation,
made by Christodoulou and Ruffini~\cite{ch,chr}: the horizon area
of a nonextremal black hole behaves in a sense as an adiabatic
invariant. Of course, the quantization of an adiabatic invariant
is perfectly natural, in accordance with the correspondence
principle.

One more conjecture made in~\cite{bek} is that the spectrum of a
quantized horizon area is equidistant. The argument therein was
that a periodic system is quantized by equating its adiabatic
invariant to $2\pi \hbar n$, $n=0,\, 1\, 2\, ...$.

Later it was pointed out by Bekenstein~\cite{be1} that the
classical adiabatic invariance does not guarantee by itself the
equidistance of the spectrum, at least because any function of an
adiabatic invariant is itself an adiabatic invariant. However, up
to now articles on the subject abound in assertions that the form
\beq\label{eq}
A = \be\, l_p^2 n, \quad n= 1,\, 2,\, ...
\eeq
for the horizon area spectrum\footnote{Here and below $l_p^2
=\hbar k/c^3$ is the Planck length squared, $l_p = 1.6 \cdot
10^{-33}$ cm, $\,k$ is the Newton gravitational constant; $\;\be$
is here some numerical factor.} is dictated by the respectable
correspondence principle. The list of these references is too
lengthy to be presented here.

Let us consider an instructive example of the situation when a
nonequidistant spectrum arises in spite of the classical adiabatic
invariance. We start with a classical spherical top of an angular
momentum $\bJ$. Of course, the $z$-projection $J_z$ of $\bJ$ is an
adiabatic invariant. If the $z$-axis is chosen along $\bJ$, the
value of $J_z$ is maximum, $J$, or $\hbar j$ in the quantum case.
The classical angular momentum squared $J^2$ is also an adiabatic
invariant, with eigenvalues $\hbar^2 j(j+1)$ when quantized. Let
us try now to use the operator $\hat{J}^2$ for the area
quantization in quite natural units of $l^2_p$. For the horizon
area $A$ to be finite in the classical limit, the power of the
quantum number $j$ in the result for $j \gg 1$ should be the same
as that of $\hbar$ in $l^2_p$~\cite{kh}. With $l^2_p \sim \hbar$,
we arrive in this way at
\[
A \sim l^2_p \sqrt{j(j+1)}.
\]
Since $\sqrt{j(j+1)} \to j+ 1/2$ for $j \gg 1$, we have come back
again to the equidistant spectrum in the classical limit. However,
the equidistance can be avoided in the following way. Let us
assume that the horizon area consists of sites with area on the
order of $l^2_p$, and ascribe to each site $i$ its own quantum
number $j_i$ and the contribution $\sqrt{j_i(j_i+1)}$ to the area.
Then the above formula changes to
\beq\label{sp}
A \sim l^2_p \sum_i \sqrt{j_i(j_i+1)}
\eeq
(in fact, this formula for a quantized area arises really as a
special case in loop quantum gravity, see below). Of course, to
retain a finite classical limit for $A$, we should require that
$\sum_i \sqrt{j_i(j_i+1)} \gg 1$. However, any of $j_i$ can be
well comparable with unity. So, in spite of the adiabatic
invariance of $A$, its quantum spectrum (\ref{sp}) is not
equidistant, though of course discrete.

One more quite popular argument in favour of the equidistant
spectrum (\ref{eq}) is as follows \cite{be1,bem,bemu}. On the one
hand, the entropy $S$ of an horizon is related to its area $A$
through the Bekenstein-Hawking relation
\beq\label{BH}
A = 4 l_p^2 S.
\eeq
On the other hand, the entropy is nothing but $\ln g(n)$ where the
statistical weight $g(n)$ of any quantum state $n$ is an integer.
In~\cite{be1,bem, bemu} the requirement of integer $g(n)$ is taken
literally, and results after simple reasoning not only in the
equidistant spectrum (\ref{eq}), but also in the following allowed
values for the numerical factor $\be$ in this spectrum:
\[
\be = 4 \ln k, \quad k=2,\, 3, \, ...\, .
\]

Let us imagine however that with some model for $S$ one obtains
for $g(n)$, instead of an integer value $K$, a noninteger one
$K+\de\,,\;\; $ $0 < \de <1$. Then, the entropy will be
\[
S = \ln(K+\de) = \ln K + \de/K.
\]
Now, the typical value of the black hole entropy $S = \ln K = A/4
l_p^2$ is huge, something like $10^{76}$. So, the correction
$\de/K$ is absolutely negligible as compared to $S = \ln K$.
Moreover, it is far below any conceivable accuracy of a
description of entropy. Therefore, this correction can be safely
omitted and forgotten. As usual for macroscopic objects, the fact
that the statistical weight is an integer has no consequences for
the entropy.

Thus, contrary to the popular belief, the equidistance of the
spectrum for the horizon area does not follow from the
correspondence principle and/or from general statistical
arguments.\\

{\bf 2.} It does not mean however that any model leading to an
equidistant spectrum for the quantized horizon area should be
automatically rejected. Quite simple and elegant version of such a
model, so called ``it from bit'', for a Schwarzschild black hole
was formulated by Wheeler~\cite{whe}. The assumption is that the
horizon surface consists  of $\nu$ patches, each of them supplied
with an ``angular momentum'' quantum number $j$ with two possible
projections $\pm 1/2$. The total number $K$ of degenerate quantum
states of this system is
\beq\label{K1/2}
K=2^\nu.
\eeq
Then the entropy of the black hole is
\beq\label{S1/2}
S_{1/2}= \ln K= \nu \ln 2.
\eeq
And finally, with the Bekenstein-Hawking relation (\ref{BH}) one
obtains for the area spectrum the following equidistant formula:
\beq\label{A1/2}
A_{1/2} = 4 \ln 2 \;l_p^2 \;\nu.
\eeq
This model of a quantized Schwarzschild black hole looks by itself
flawless.

Later this result was derived in Ref.~\cite{asht} in the framework
of loop quantum gravity (LQG)~[10-14]. We discuss below whether
the ``it from bit'' picture, if considered as a special case of
the area quantization in LQG, can be reconciled with the
holographic bound~[15-17].

More generally, a quantized surface in LQG is described as
follows. One ascribes to it a set of punctures. Each puncture is
supplied with two integer or half-integer ``angular momenta''
$j^u$ and $j^d$:
\beq\label{j}
j^u,\, j^d= 0, 1/2, 1, 3/2, ...\;.
\eeq
$j^u$ and $j^d$ are related to edges directed up and down the
normal to the surface, respectively, and add up into an angular
momentum $j^{ud}$:
\beq\label{ud}
{\bf j}^{ud}= {\bf j}^{u} + {\bf j}^{d}; \quad |j^{u}-j^{d}|\leq
j^{ud} \leq j^{u}+j^{d}.
\eeq
The area of a surface is
\beq\label{Aj}
A =\beta\, l_p^2 \sum_i \sqrt{2 j^u_i(j^u_i+1)+ 2j^d_i(j^d_i+1)-
j^{ud}_i(j^{ud}_i+1)}\;.
\eeq
The overall numerical factor $\beta$ in (\ref{Aj}) cannot be
determined without an additional physical input. This ambiguity
originates from a free (so-called Barbero-Immirzi) parameter
\cite{bar,imm} which corresponds to a family of inequivalent
quantum theories, all of them being viable without such an input.

The result (\ref{A1/2}) was obtained in~\cite{asht} under an
additional condition that the gravitational field on the horizon
is described by the $U(1)$ Chern-Simons theory. Formula
(\ref{A1/2}) is a special case of general one (\ref{Aj}) when all
$j^d$ vanish and all $j^u$ equal $1/2$ (or vice versa). As to the
overall factor $\be$, its value here is \footnote{The common
convention for the numerical factor in formula (\ref{Aj}) is
$8\pi\be$; with it the parameter $\be$ is smaller than ours by
factor $8\pi$.}
\beq\label{be}
\beta = \frac{8\ln 2}{\sqrt{3}}\,.
\eeq

Let us turn now to the holographic bound~[15-17]. According to it,
the entropy $S$ of any spherically symmetric system confined
inside a sphere of area $A$ is bounded as follows:
\beq\label{hb}
S \leq \frac{A}{4 l_p^2}\,,
\eeq
with the equality attained only for a system which is a black
hole.

A simple intuitive argument confirming this bound is as
follows~\cite{sus}. Let us allow the discussed system to collapse
into a black hole. During the collapse the entropy increases from
$S$ to $S_{bh}$, and the resulting horizon area $A_{bh}$ is
certainly smaller than the initial confining one $A$. Now, with
the account for the Bekenstein-Hawking relation (\ref{BH}) for a
black hole we arrive, through the obvious chain of (in)equalities
\[
S \leq S_{bh} = \frac{A_{bh}}{4 l_p^2} \leq \frac{A}{4 l_p^2}\,,
\]
at the discussed bound (\ref{hb}).

The result (\ref{hb}) can be formulated otherwise. Among the
spherical surfaces of a given area, it is the surface of a black
hole horizon that has the largest entropy.

On the other hand, it is only natural that the entropy of an
eternal black hole in equilibrium is maximum. This was used by Vaz
and Witten~\cite{vaz} in a model of the quantum black hole as
originating from a dust collapse. Then the idea was employed by
us~\cite{k,kk} in the problem of quantizing the horizon of a black
hole in LQG. In particular, the coefficient $\be$ was calculated
in Ref.~\cite{kk} in the case when the area of a black hole
horizon is given by the general formula (\ref{Aj}) of LQG, as well
as under some more special assumptions on the values of $j^u$,
$j^d$, $j^{ud}$. Moreover, it was demonstrated in Ref.~\cite{kk}
for a rather general class of the horizon quantization schemes
that it is the maximum entropy of a quantized surface which is
proportional to its area.

Let us sketch the proof of this result (for more technical details
see~\cite{kk}). We consider here and below in the present paper
the microcanonical entropy $S$ of a surface (though with fixed
area instead of fixed energy). It is defined as the logarithm of
the number of states of this surface with a fixed area $A$, i. e.
with a fixed sum
\beq\label{N}
N\;=\; \sum_i \sqrt{2 j^u_i(j^u_i+1)+ 2j^d_i(j^d_i+1)-
j^{ud}_i(j^{ud}_i+1)}\;.
\eeq
Let $\nu_{im}$ be the number of punctures with a given set of
momenta $j^u_i$, $j^d_i$, $j^{ud}_i$, and a given projection $m$
of $j^{ud}_i$. The total number of punctures is
\[
\nu = \sum_{im} \nu_{im}.
\]
We will assume that the edges with the same set of the quantum
numbers $im$ (i. e. with the same $j^u_i$, $j^d_i$, $j^{ud}_i$,
and $m$) are indistinguishable, so that interchanging them does
not result in new states. All other permutations, those among the
edges with differing $im$, do create new states, so that such
edges, with differing $im$, are distinguishable \footnote{Let us
note that the ``it from bit'' values (\ref{K1/2}) and (\ref{S1/2})
for the number of states and entropy, also follow from this
assumption. Indeed, let $\nu$ be the total number of patches with
$j=1/2$, and let $\;\nu_+$ and $\nu_-=\nu - \nu_+$ patches have
the projections $+1/2$ and $-1/2$, respectively. Then, the number
of the corresponding states is obviously
\[
\frac{\nu!}{\nu_+!\; (\nu-\nu_+)!}\,,
\]
and the total number of states is
\[
K=\sum_{\nu_+=0}^{\nu}\frac{\nu!}{\nu_+!\; (\nu-\nu_+)!}=2^{\nu},
\]
in agreement with (\ref{K1/2}).}. Then, the entropy is
\beq\label{en1}
S=\ln\left[\nu\,!\,\prod_{im}\,\frac{1}{\nu_{im}\,!} \right].
\eeq
The structure of expressions (\ref{Aj}) and (\ref{en1}) is so
different that in a general case the entropy certainly cannot be
proportional to the area. However, this is the case for the
maximum entropy in the classical limit.

By combinatorial reasons, it is natural to expect that the
absolute maximum of entropy is reached when all values of quantum
numbers $j_i^{u,d,ud}$ are present. We assume also that in the
classical limit the typical values of puncture numbers $\nu_{im}$
are large. Then, with the Stirling formula for factorials,
expression (\ref{en1}) transforms to
\beq\label{en2}
S= \sum_{im} \nu_{im} \times \ln \left(\sum_{i^{\prime}m^{\prime}}
\nu_{i^{\prime}m^{\prime}}\right)- \sum_{im} \nu_{im} \,\ln
\nu_{im} \,.
\eeq

We are looking for the extremum of expression (\ref{en2}) under
the condition
\beq\label{con}
N=\sum_i \nu_{im}\,r_i\, = {\rm const},
\eeq
where each partial contribution $ r_i=\sqrt{2 j^u_i(j^u_i+1)+
2j^d_i(j^d_i+1)- j^{ud}_i(j^{ud}_i+1)}$ is independent of $m$. The
problem reduces to the solution of the system of equations
\beq\label{sys}
\ln \left(\sum_{i^{\,\prime}m^{\,\prime}}
\nu_{i^{\,\prime}m^{\,\prime}}\right) - \ln \nu_{im} = \mu r_i\,,
\eeq
or
\beq\label{nu}
\nu_{im} = e^{- \mu r_i}\,\sum_{i^{\,\prime}m^{\,\prime}}
\nu_{i^{\,\prime}m^{\,\prime}}\,=\nu\,e^{- \mu r_i}.
\eeq
Here $\mu$ is the Lagrange multiplier
for the constraining relation (\ref{con}). Summing expressions
(\ref{nu}) over $i,m$, we arrive at the equation for $\mu$:
\beq\label{equ}
\sum_{im} \, e^{- \mu r_i}= \sum_i g_i\, e^{- \mu r_i}= 1;
\eeq
the statistical weight $g_i = 2 j^{ud}_i + 1$ of a puncture arises
here since $r_i$ are independent of $m$. On the other hand, when
multiplying equation (\ref{sys}) by $\nu_{im}$ and summing over
$i,m$, we arrive with the constraint (\ref{con}) at the following
result for the maximum entropy for a given value of the sum~$N$,
or the black hole area $A$:
\beq\label{enf}
S_{\rm max}= \mu \,N\,=\,\frac{\mu}{\be l^2_p}\,A.
\eeq

One more curious feature of the obtained picture is worth noting:
it gives a sort of the Boltzmann distribution for the occupation
numbers (see (\ref{nu})). In this distribution, the partial
contributions $r_i$ to the area are analogues of energies, and the
Lagrange multiplier $\mu$ corresponds (up to a factor) to the
inverse temperature.

It should be emphasized that relation (\ref{enf}) is true not only
in LQG, but applies to a more general class of approaches to the
quantization of surfaces. The following assumption is really
necessary here: the surface should consist of patches of different
sorts, so that there are $\nu_{im}$ patches of each sort $i,m$,
with a generalized effective quantum number $r_i$, and a
statistical weight $g_i$. As necessary is the above assumption on
the distinguishability of the patches.

Thus, it is the maximum entropy of a surface which is proportional
in the classical limit to its area. This proportionality certainly
takes place for a classical black hole. And this is one more
strong argument in favour of the assumption that the black hole
entropy is maximum.

Let us come back now to the result of Ref.~\cite{asht}. If one
assumes that the value (\ref{be}) of the parameter $\be$ is the
universal one (i. e. it is not special to black holes, but refers
to any quantized spherical surface), then the value (\ref{S1/2})
is not the maximum one in LQG for a surface of the area
(\ref{A1/2}). This looks quite natural: with the transition from
the unique choice made in Ref.~\cite{asht}, $j^{u(d)}=1/2$,
$j^{d(u)}=0$, to more extended and rich one, the number of the
degenerate quantum states should, generally speaking, increase.
And together with this number, its logarithm, which is the entropy
of a quantized surface, increases as well.

We start the proof of the above statement with rewriting formula
(\ref{S1/2}) as follows:
\beq\label{S'}
S_{1/2}=\, \ln 2\,\sqrt{\frac{4}{3}}\,N\, = \,0.80\, N; \quad N=
\sqrt{\frac{3}{4}}\,\nu\,.
\eeq
From now on, we consider this value of $N$ as fixed one.

Let us start with a relatively simple example when $j^{d(u)}=0$,
so that the general formula (\ref{Aj}) for a surface area reduces
to
\beq\label{A1}
A =\be\, l_p^2 \sum_i \sqrt{j_i(j_i+1)}\;=\be\, l_p^2
\sum_{j=1/2}^{\infty}\sqrt{j(j+1)}\;\nu_j\,, \quad j=j^{u(d)}
\eeq
(and coincides with our naive model~(\ref{sp})). We will find the
maximum entropy of such a surface for the fixed value of
\beq\label{N1}
N=\sum_{j=1/2}^{\infty}\sqrt{j(j+1)}\;\nu_j\,,
\eeq
that should be equal to the ``it from bit'' one, $\nu\sqrt{3/4}$.
Here the statistical weight of a puncture with the quantum number
$j$ is $g_j= 2j + 1$, and equation (\ref{equ}) can be rewritten as
\beq\label{eq1}
\sum_{p=1}^{\infty}(p+1)\,z^{\sqrt{p(p+2)}}=1, \quad p=2j, \quad
z=e^{-\mu/2}.
\eeq
Its solution is $\mu=-2\ln z = 1.722$~\cite{kk}, and the maximum
entropy in this case
\beq\label{s1}
S_{\rm max, 1}=\,1.72\,N
\eeq
exceeds the result (\ref{S'}).

As expected, in the general case, with $N$ given by formula
(\ref{N}) with all values of $j^u_i$, $j^d_i$, $j^{ud}_i$ allowed
and $g_i= 2j^{ud}_i +1$, the maximum entropy is even
larger~\cite{kk}
\beq\label{s3}
S_{\rm max}=\,3.12\,N.
\eeq

Thus, the conflict is obvious between the holographic bound and
the result (\ref{S'}), as found within the LQG approach
of~\cite{asht}.

One might try to avoid the conflict by assuming that the value
(\ref{be}) for the Barbero-Immirzi parameter $\be$ is special for
black holes only, while for other quantized surfaces $\be$ is
smaller. However, such a way out would be unattractive and
unnatural.

{\bf 3.} We come back now to the essential assumption made in the
previous section: the edges with the same set of the quantum
numbers $im$ are identical, the edges with differing $im$ are
distinguishable. In principle, one might try to modify this
assumption of partial distinguishability of edges in two opposite
ways.

One possibility, which might look quite appealing, is that of
complete indistinguishability of edges. It means that no
permutation of any edges results in new states. To simplify the
discussion, let us confine here and below to expression (\ref{A1})
for the horizon area, instead of the most general one (\ref{Aj}).
Then, the total number of angular momentum states created by
$\nu_j= \sum_m \nu_{jm}$ indistinguishable edges of a given $j$
with all $2j+1$ projections allowed, from $-j$ to $j$, is
\footnote{Perhaps, the simplest derivation of this formula is as
follows. We are looking here effectively for the number of ways of
distributing $\nu_j$ identical balls into $2j+1$ boxes. Then, the
line of reasoning presented in \cite{lls}, \S 54, results in
formula (\ref{id}). I am grateful to V.F. Dmitriev for bringing to
my attention that formula (\ref{id}) can be derived in this
simpleminded way.}
\beq\label{id}
K_j = \,\frac{(\nu_j+2j)!}{\nu_j!\; (2j)!}\,.
\eeq

Those partial contributions $s_j=\ln K_j$ to the black hole
entropy $S=\sum_j s_j$ that can potentially dominate the
numerically large entropy, may correspond to the three cases: $j
\ll \nu_j\,$, $j \gg \nu_j\,$, and $j \sim \nu_j \gg 1$. These
contributions are as follows:\\

\begin{tabular}[h]{ll}
\vspace{3mm} $j \ll \nu_j\,$, & $s_j \approx 2j\ln \nu_j\,$; \\
 \vspace{3mm}
$j \gg \nu_j\,$,  & $s_j \approx \nu_j\ln j$; \\
 \vspace{3mm}
$j \sim \nu_j \gg 1$,  & $s_j \sim 4j\ln 2$.\\
\end{tabular}

In all the three cases the partial contributions to the entropy
$S$ are much smaller parametrically than the corresponding
contributions
\\

 \hspace{3mm} $a_j \sim j \nu_j$\\

\noindent to the area $A=\sum_j a_j$. Thus, in all these cases $S
\ll A$, so that with indistinguishable edges of the same $j$, one
cannot make the entropy of a black hole proportional to its area.
It was pointed out earlier in Refs. \cite{aps,khr}.

Let us consider now the last conceivable option, that of
completely distinguishable edges. In this case the total number of
states is just $K = \nu\,!$, instead of (\ref{en1}), with the
microcanonical entropy $S=\nu \ln \nu$. In principle, this entropy
can be made proportional to the black hole area $A$. The model
(though not looking natural) could be as follows. We choose a
large quantum number $J  \gg 1$, and assume that the horizon area
$A$ is saturated by the edges with $j$ in the interval $J~<j~<2J$,
and with ``occupation numbers'' $\nu_j \sim \ln J$. Then, the
estimates both for $S$ and $A$ are $\sim J\ln J$, and the
proportionality between the entropy and the area can be attained.

However, though under the assumption of complete
distinguishability the entropy can be proportional to the area,
the {\it maximum} entropy for a given area is much larger than the
area itself. Obviously, here the maximum entropy for fixed $A \sim
\sum_j \sqrt{j(j+1)}\; \nu_j$ is attained with all $j$'s being as
small as possible, say, $1/2$ or 1. Then, in the classical limit
$\nu \gg 1$, the entropy of a black hole grows faster than its
area, $A \sim \nu$, while $S = \nu \ln \nu \sim A\ln A$. Thus, the
assumption of complete distinguishability is in conflict with the
holographic bound, and therefore should be discarded.

There is no disagreement between this our conclusion and that of
Refs. \cite{aps,pol,gs}: what is called complete
distinguishability therein corresponds to our partial
distinguishability.

\begin{center}***\end{center}
I am grateful to V.F. Dmitriev, V.M. Khatsymovsky, and G.Yu. Ruban
for discussions. The investigation was supported in part by the
Russian Foundation for Basic Research through Grant No.
03-02-17612.


\begin{thebibliography}{99}
\bibitem{bek} J.D. Bekenstein, Lett. Nuovo Cimento 11 (1974) 467.
\bibitem{ch} D. Christodoulu, Phys. Rev. Lett. 25 (1970) 1596.
\bibitem{chr} D. Christodoulu, R. Ruffini, Phys. Rev. D 4 (1971) 3552.
\bibitem{be1} J.D. Bekenstein, in \textit{Proceedings of the
Eighth Marcel Grossmann Meeting\\ on General Relativity}, ed. by
Tsvi Piran and Remo Ruffini (World Scientific,\\ Singapore, 1999)
92; gr-qc/9710076.
\bibitem{kh} I.B. Khriplovich, Phys. Lett. B431 (1998) 19;
gr-qc/9804004.
\bibitem{bem} J.D. Bekenstein, V.F. Mukhanov, Phys. Lett. B360 (1995) 7;
gr-qc/9505012.
\bibitem{bemu} J.D. Bekenstein, V.F. Mukhanov, in \textit{Sixth
Moscow Quantum Gravity Seminar}, ed. by~V.A. Berezin, V.A.
Rubakov, and D.V. Semicoz (World Scientific, Singapore, 1997).
\bibitem{whe} J.A. Wheeler, in \textit{Sakharov Memorial Lectures
in Physics,} Moscow, 1991, vol. 2, 751.
\bibitem{asht} A. Ashtekar, J. Baez, A. Corichi, K. Krasnov,
Phys. Rev. Lett. 80 (1998) 904;\\ gr-qc/9710007.
\bibitem{rov} C. Rovelli, L. Smolin, Nucl. Phys. B 442 (1995) 593;
erratum, ibid. B 456 (1995) 753; gr-qc/9411005.
\bibitem{ash} A. Ashtekar, J. Lewandowski, Class. Quantum Grav. 14 (1997) 55;
gr-qc/9602046.
\bibitem{lol} R. Loll, Phys. Rev. Lett. 75 (1995) 3048;
gr-qc/9506014;\\ R. Loll, Nucl. Phys. B 460 (1996) 143;
gr-qc/9511030.
\bibitem{dep} R. De Pietri, C. Rovelli, Phys. Rev. D 54 (1996) 2664;
gr-qc/9602023.
\bibitem{leh} S. Frittelli, L. Lehner, C. Rovelli, Class. Quantum Grav. 13
(1996) 2921; gr-qc/9608043.
\bibitem{bek1} J.D. Bekenstein, Phys. Rev. D 23 (1981) 287.
\bibitem{tho} G. 't Hooft, in \textit{Salam Festschrift,} Singapore, 1993;
gr-qc/9310026.
\bibitem{sus} L. Susskind, J. Math. Phys. 36 (1995) 6377;\\
gr-qc/9710007.
\bibitem{bar} J.F. Barbero G., Phys. Rev. D 51 (1995) 5507;
gr-qc/9410014.
\bibitem{imm} G. Immirzi, Class. Quantum Grav. 14 (1997) L177; gr-qc/9701052.
\bibitem{vaz} C. Vaz, L. Witten, Phys. Rev. D 64 (2001) 084005; gr-qc/0104017.
\bibitem{k} I.B. Khriplovich, Phys. Lett. B 537 (2002) 125; gr-qc/0109092.
\bibitem{kk}  R.V. Korkin, I.B. Khriplovich, Zh. Eksp. Teor. Fiz. 122
(2002) 1\newline $[\,$Sov. Phys. JETP 95 (2002) 1$\,]$;
gr-qc/0112074.
\bibitem{aps} A. Alekseev, A.P. Polychronakos, M. Smedback, Phys. Lett. B 574 (2003)
296;\\ hep-th/0004036.
\bibitem{khr} I.B. Khriplovich, in {\it Proceedings of the St. Petersburg School
of Physics},\\ St. Petersburg, 2002; gr-qc/0210108.
\bibitem{pol} A.P. Polychronakos, Phys. Rev. D 69 (2004) 044010;
hep-th/0304135.
\bibitem{gs} G. Gour, V. Suneeta, gr-qc/0401110.
\bibitem{lls} L.D. Landau, E.M. Lifshitz, {\it  Statistical Physics}, (Pergamon
Press, London 1958).

\end{thebibliography}
\end{document}